\documentclass[fleqn,10pt]{wlscirep}
\usepackage[utf8]{inputenc}
\usepackage[T1]{fontenc}

\usepackage{svg}
\usepackage{soul}
\usepackage{subfigure}
\usepackage{amssymb, bm}
\usepackage{pgfplots}
\usepackage{multirow}
\usepackage{amsfonts}
\usepackage{comment}
\usepackage{booktabs}
\usepackage{siunitx}
\usepackage{xcolor}
\usepackage{float}
\usepackage{authblk}
\usepackage[ruled,vlined, noend]{algorithm2e}
\usepackage{caption}
\usepackage{amsmath}
\usepackage[T1]{fontenc}
\usepackage[utf8]{inputenc}

\definecolor{rev}{rgb}{1,1,1}
\definecolor{revrev}{rgb}{1,1,1}
\usepackage{tikz}
\usepackage[draft]{todonotes}
\usepackage{amsopn}
\usepackage{mdframed}
\usepackage[pagewise]{lineno}

\usetikzlibrary{arrows.meta}

\usepackage[draft]{todonotes}
\usepackage{amsopn}
\PassOptionsToPackage{hyphens}{url}\usepackage{hyperref}

\usepackage[normalem]{ulem}

\title{Modelling virus spreading in ride-pooling networks \\ 
\small Published in \emph{Scientific Reports 11 (1), 1-11 }\url{https://doi.org/10.1038/s41598-021-86704-2}}
\author[1,2,*]{Rafa\l{} Kucharski}
\author[1]{Oded Cats} 
\author[3]{Julian Sienkiewicz}
\affil[1]{\small Department of Transport \& Planning, Delft University of Technology, the Netherlands}
\affil[2]{\small Department of Transport Systems, Cracow University of Technology, Poland}
\affil[3]{\small Faculty of Physics, Warsaw University of Technology, Poland}

\affil[*]{\small corresponding author: Rafa\l{} Kucharski, r.m.kucharski@tudelft.nl}


\begin{abstract}
Urban mobility needs alternative sustainable travel modes to keep our pandemic cities in motion.  Ride-pooling, where a single vehicle is shared by more than one traveller, is not only appealing for mobility platforms and their travellers, but also for promoting the sustainability of urban mobility systems. Yet, the potential of ride-pooling rides to serve as a safe and effective alternative given the personal and public health risks considerations associated with the COVID-19 pandemic is hitherto unknown.

To answer this, we combine epidemiological and behavioural shareability models to examine spreading among ride-pooling travellers, with an application for Amsterdam. Findings are at first sight devastating, with only few initially infected travellers needed to spread the virus to hundreds of ride-pooling users. Without intervention, ride-pooling system may substantially contribute to virus spreading. Notwithstanding, we identify an effective control measure allowing to halt the spreading before the outbreaks (at 50 instead of 800 infections) without sacrificing the efficiency achieved by pooling. 
Fixed matches among co-travellers disconnect the otherwise dense contact network, encapsulating the virus in small communities and preventing the outbreaks.

\end{abstract}

\begin{document}

\flushbottom
\maketitle
\thispagestyle{empty}
%\linenumbers
\par

In the era of widespread concerns about personal safety and exposure to virus transmission, urban mobility faces an unprecedented challenge \cite{Acuto2020, He2020,muller2020mobility}. While mass transit, a crowded backbone of pre-pandemic megacities' mobility systems, is under societal pressure due to health concerns related to its potential role in virus spreading\cite{UITP,gkiotsalitis2020optimal}, people search for other travel alternatives that reduce one’s exposure.
The natural reaction of risk-averse travellers is to opt for individual transport modes, such as private cars\cite{tirachini2020covid}, which can be devastating for the sustainability of pandemic urban mobility systems \cite{guerriero2020health}. To counteract this, we explore whether shared mobility may offer an attractive alternative by efficiently serving travel demand using a shared fleet while allowing users to avoid the crowd.
\textbf{Ride-pooling}\cite{alonso2017demand, kucharski2020exact}, available via two-sided mobility platforms (such as \texttt{UberPool} and \texttt{Lyft}), has recently emerged as a travel alternative in cities worldwide and gained attention or researchers studying both specific systems (like Singapore\cite{Yang2020} and New York City\cite{Riascos2020}) as well as uncovering universal scaling laws governing cities\cite{Tachet2017,Chen2018, santi2014quantifying} which in turn allow for generalizing the results to any urban system.
Travellers, requesting rides are offered a pooled ride, where they share a single vehicle with co-travellers riding in a similar direction. 
%Both pick-up and travel times may deviate from the desired or minimal ones, since the vehicle needs to meet the requirements of all pooled travel requests. 
%The service provider needs to compensate this inconvenience by offering a lower fare. 
Despite being perceived by policymakers as a solution for improving mobility and sustainability by leveraging on the platform economy revolution, the COVID pandemic led to safety concerns among sharing travellers (worried about their health), policymakers (concerned about public health and epidemic outbreak) and operators (uncertain about the future of their business).

While preliminary findings on COVID-19 \cite{liu2020secondary} suggest transmission taking place in proximity (e.g. among co-travellers within the same vehicle), evidence on how and if the virus transmits beyond a single vehicle is lacking. Indeed, the potential of shared rides to serve as an alternative, in-between the mass transit (where - perceived or real - virus exposure may be high) and private cars (which generate negative externalities) remains largely unknown. 
Will the random infected passenger spread the virus across a large number of travellers across the network, or will it be encapsulated and thus confined to a distinct community? How many other travellers will get infected and how will the epidemiological process evolve? Finally, can we mitigate it by effective control and design measures and thus introduce it to policymakers as a safe alternative? Such questions are valid as COVID-19 pandemic constantly challenge our current policies\cite{DellaRossa2020, LiuZhu2020} calling for a long-term preparedness\cite{editorial}.

The propagation of different types of epidemics (biological and social) has long been a playground of network science community (summed up in the seminal work of Pastor-Satorras et al.\cite{pastor2015epidemic} and reaching as far as to propose the idea of ``physics of vaccination''\cite{wang2016}). Addressing also mobility network (e.g. public transport) structures \cite{Sienkiewicz2005,Barthelemy2011} with their complex topology and temporal evolution studies\cite{Gallotti2015}. Recent COVID-19 propagation studies either follow a coarse-grained level of aggregated cases\cite{Aleta2020,Chinazzi2020,Kraemer2020}, adopt purely synthetic network structures\cite{Azizi2020,xue2020} or lack emerging mobility modes (like ride-pooling) in the picture\cite{muller2020mobility, tirachini2020covid}. 
This study brings to the front the network evolution (crucial in the context of epidemic spreading\cite{Ciaperoni2020,Cacciapaglia2020}) and couples it with empirical, behaviour driven contact network \cite{kucharski2020exact}, specific to ride-pooling, yielding a simulation framework (see Methods for details) capable to provide rich and realistic insights into possible epidemic outbreaks specific to ride-pooling networks.

\begin{figure}
\centering
\includegraphics[width=0.99\linewidth]{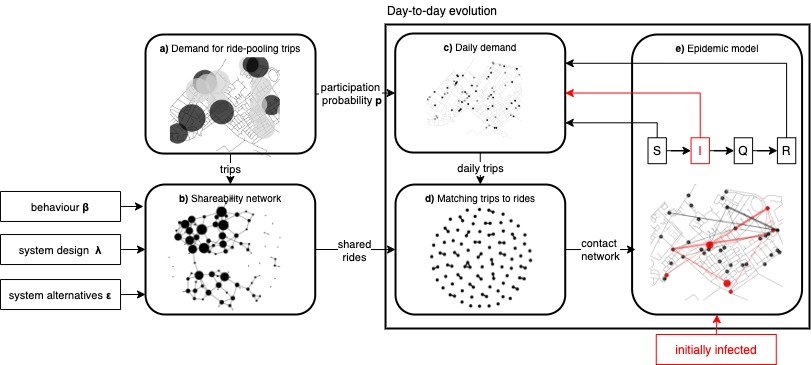}
\caption{Methodology at glance: We consider travel demand for ride-pooling trips (a), for which we compute a shareability network (b) with a given behavioural parameters $\beta$, system design $\lambda$ and alternatives' attractiveness $\epsilon$. We simulate the day-to-day evolution of spreading until the virus is halted. Each day we obtain the daily demand (c), consisting of those who want and can travel (decided to travel with probability $p$ and are not quarantined). Daily trip demand is optimally assigned to shared rides, which forms the contact network (d) on which virus spreading is then modelled (e). Starting from initially infected travellers, each day we simulate epidemic transitions: susceptible travellers are infected by infected co-travellers who quarantine after 7 days and return immune to the system after 14 days.}
\label{fig:stream}
\end{figure}

We model the evolution of virus spreading in ride-pooling systems through an extensive set of experiments with demand sampled from actual mobility patterns of afternoon commuters in Amsterdam. The underlying {\bf shareability network} (see Methods for details how such network is set up) is the outcome of travellers' willingness to share, which depends on whether they are sufficiently close to each other in terms of induced detour (compatibility of origins and destinations) and delay (compatibility of departure times) compared to a private ride-hailing. The resulting dynamic, time-dependent contact network \cite{volz2009epidemic} is subject to day-to-day variations as well as the results of the iterative {\bf SIQR epidemiological model} \cite{Feng1995} (see Methods section for explanation). We examine the resulting spreading process, i.e. number of infections along with its temporal and spatial evolution. To instantly show the methodology at glance, Fig. \ref{fig:stream} presents all steps of our framework and their inter-dependencies, further detailed in Methods section.

Findings from our extensive simulation study are on first sight devastating, with only few initially infected travellers needed to spread the virus to hundreds of ride-pooling users. Even under conservative assumptions where the driver is not a spreader and mobility pattern is restricted to only two trips per day, the virus makes its way to infect the majority of the giant component. Introducing natural stochasticity and non-recurrence inherent to travel demand triggers a virus transmission. Despite slow temporal evolution virus gradually makes its way towards new communities, neglecting natural spatial barriers. There seems to be no epidemic threshold and even two initial infections may trigger an outbreak and reach high transmissivity. This is a very alarming finding, suggesting that ride-pooling system without intervention may substantially contribute to virus spreading. 
Nonetheless, we identified effective control measure allowing to halt the spreading before the outbreaks. Namely, if we trade-off spontaneity of platform-based ride-pooling service and let the operator fix matches with co-travellers, radically different image appears. 
Such setting disconnects the otherwise dense contact network, containing the virus in small communities and preventing the outbreaks. 
Notably, this trade-off is not at the cost of system effectiveness, most importantly not at the cost of occupancy rate, which may remain at the original level. 
We argue that under strict demand control measures, mobility platforms may provide an appealing alternative service in-between public and private transport modes for pandemic reality.
Universal properties of ride-pooling networks\cite{santi2014quantifying,Tachet2017,Molkenthin2020} allow us to generalise our Amsterdam findings to a generic systems for which the critical mass needed to induce sharing comes along with a highly connected shareability network, whereas fixing the matches disconnects it into isolated communities.

\begin{figure}
  \centering
  \subfigure[]{\includegraphics[scale=0.26]{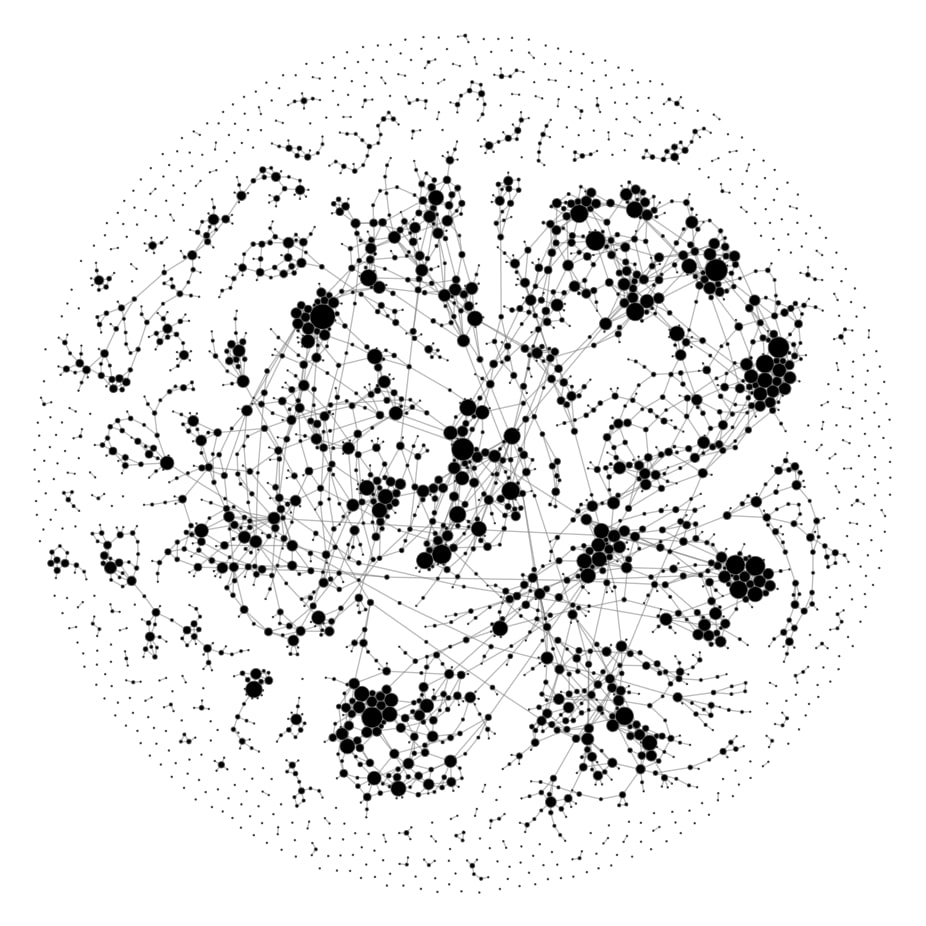}}\quad
  \subfigure[]{\includegraphics[scale=0.17]{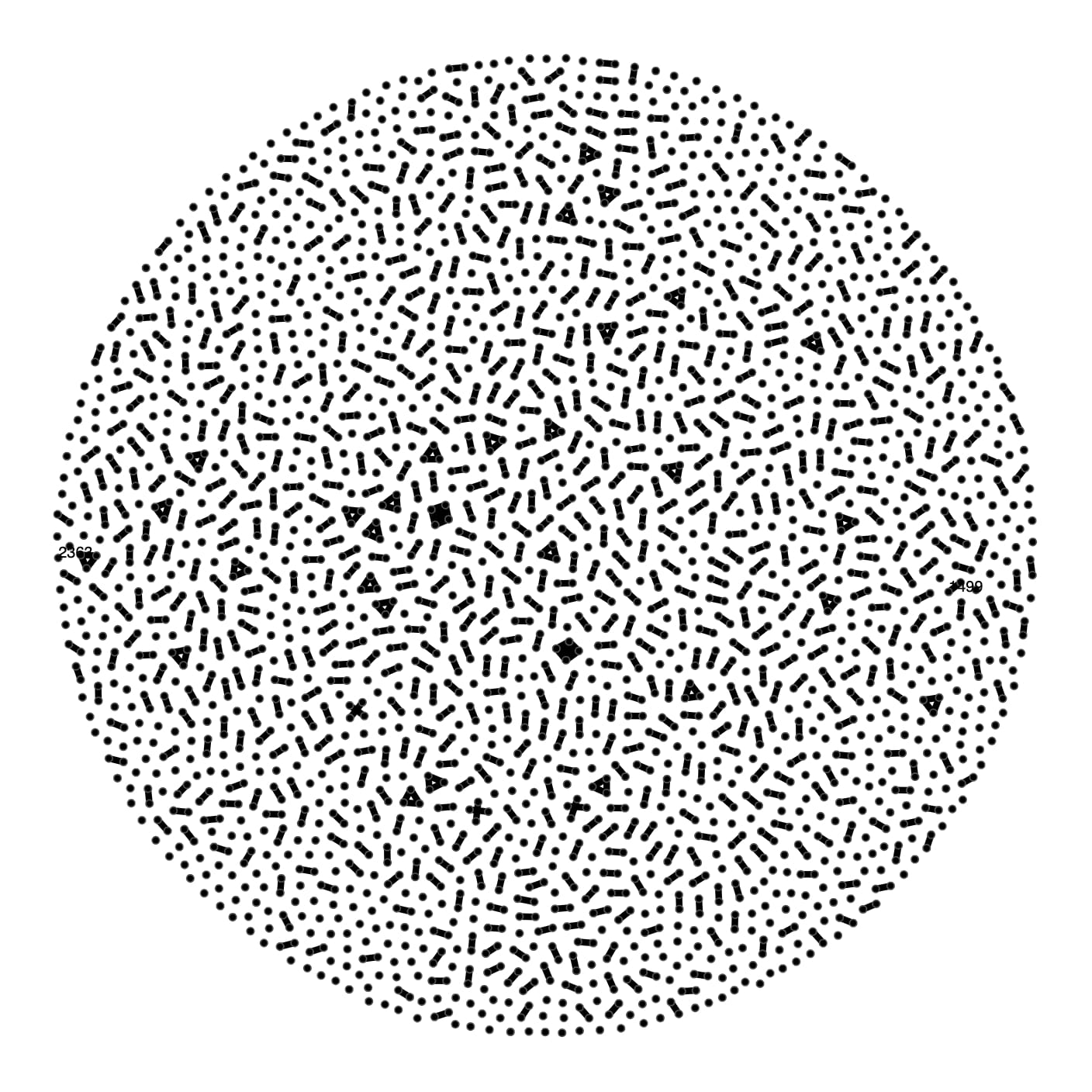}}
\caption{\small Shareability graph linking 3 200 travellers to 11 000 pooled rides feasible for them (a). Size of nodes is proportional to degree (number of travellers for shared rides and number of feasible rides for travellers). The graph structure includes a giant component and a high degree nodes, which may become a super-spreaders, as well as isolated peripheral nodes, where travellers either cannot find a feasible match or form a small, isolated communities from which virus will not outbreak. The actual matching of travellers to shared rides on a single day (b) has a substantially different structure. Here (b) nodes denote travellers, linked if they share a ride. Single dots are unmatched travellers riding alone, while lines, triangles and squares denote pooled rides of higher degree (2,3 and 4, respectively). While the potential shareability (a) is densely connected, matching on a single day (b) is disintegrated. Each pooled-ride forms an isolated community (i.e. co-travellers within a single vehicle), with a clique of size bounded with vehicle capacity (four in our case). The virus will spread within each clique but will not reach beyond it on a given day. However, infected traveller may be assigned to a new ride on successive day, resulting with virus propagation beyond the single vehicle. Networks visualized with \texttt{newtulf}\cite{aslak2019netwulf}}
\label{fig:graf}
\end{figure}

\section*{Application and Results}
To understand how the virus spreads among travellers sharing rides, we conduct a series of experiments within a realistic travel demand setting of Amsterdam. Afternoon commuters, sampled from the actual trip demand dataset\cite{arentze2004learning}, hail a ride from a mobility platform to reach their destination. 
They may opt for a shared ride if reduced trip fare will compensate for any detour and delay imposed by sharing. We consider a system where a 30\% discount is offered for sharing and we specify the private ride-hailing ride as an alternative $\epsilon$. 
We employ behavioural parameters (value-of-time and willingness-to-share $\beta$'s) in-line with recent findings \cite{alonso2020determinants, alonso2020value}
and apply \textit{ExMAS} algorithm \cite{kucharski2020exact} to reproduce a behaviourally rich shareability network connecting {3 200} travellers to {11 000} feasible shared rides (see Methods for algorithm description). 
The size of travel demand sample is such that, on one hand, the critical mass needed to induce sharing can be attained and on the other hand, it represents a relatively low demand levels reached by ride-pooling services so far \cite{LI2019330}. Notably, we model demand for shared rides which is non-stable and fluctuating from one day to the other \cite{lyon}, hence each afternoon is comprised of a slightly or significantly different pool of travellers, controlled through a {\it demand stability} parameter $p$ (i.e., the participation probability, see Methods). To allow for comparisons, while experimenting with $p$ we keep the total daily number of travellers in the system fixed (to 2000), yet we adjust the pool of passenger from which we draw them on any given day.
In our series of experiments we explore demand stability varying from $p=0.65$ (where each day we draw from the pool of $3075 = 2000/0.65$ travellers) up to $p=1$ (where the total demand is assumed constant). 
We vary the number of initially infected travellers from 2 to 20. 
To assess the impact of demand level, we conduct experiments where we gradually increase it up to 3200 travellers.
In order to account for the impact of their random location, we replicate each scenario 20 times.

\begin{figure}[ht!]
    \centering
    \includegraphics[width=0.99\textwidth]{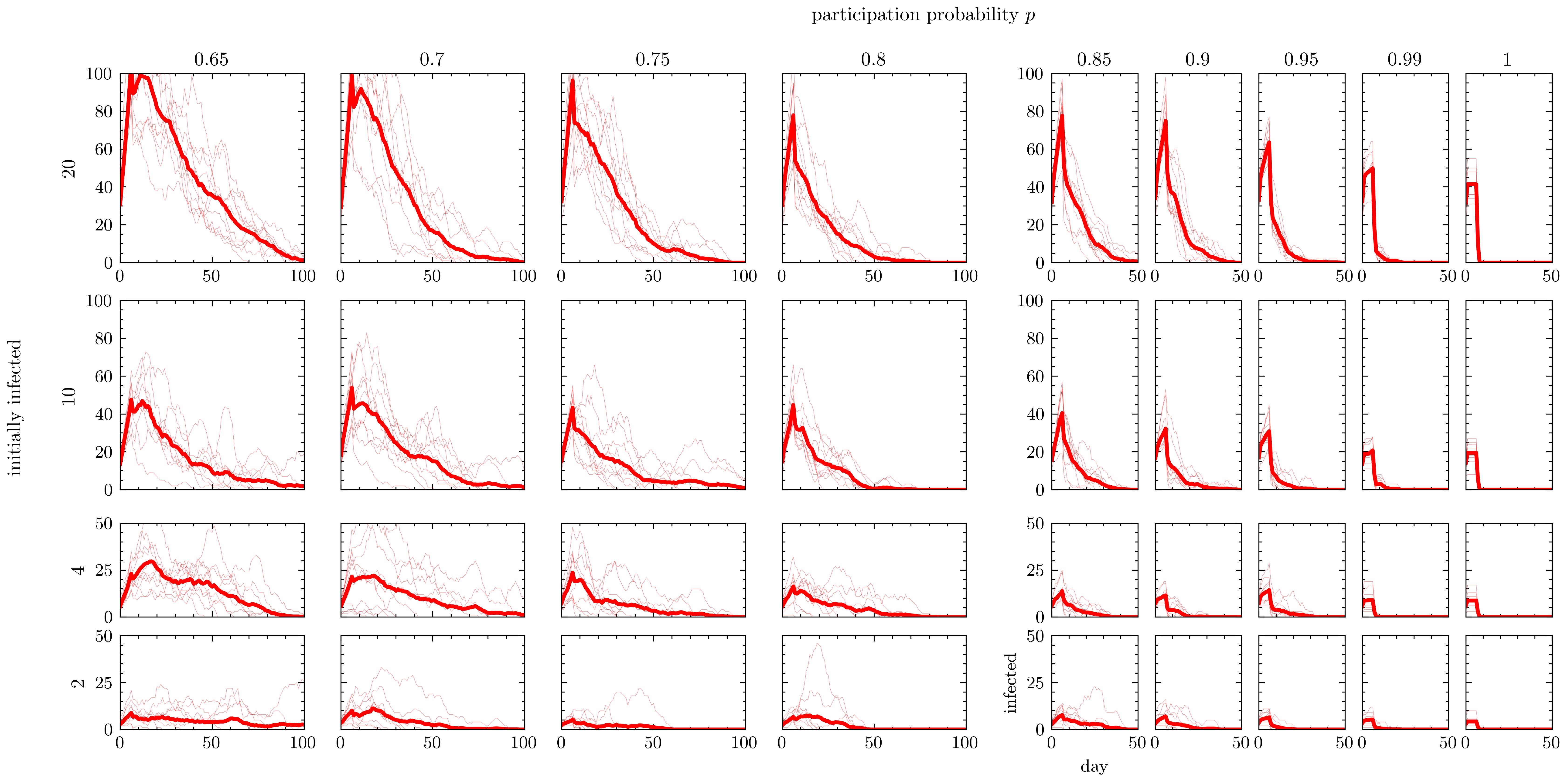}
    \caption{\small Number of infected travellers over the course of epidemic outbreaks, with various settings of initially infected (rows) and demand stability (columns), bold lines denote averages over all experiments (shown individually using thin lines). With an unstable demand (0.65), 20 initially infected always triggers transmission reaching at least 60 travellers (out of 2000) and lasting at least 60 days. Yet in most other configurations results are less stable and actual outbreaks strongly depends on the location of initially infected, revealing a strongly heterogeneous structure of the underlying contact network. In most cases we can observe a smooth, log-normal shape with a strong outbreak in the first phase, exponentially decaying in the latter phases. Mean temporal profiles of outbreaks are consistently following the trend of decreasing when the number of initial infections is lower and a demand pattern becomes more stable . For stable demand patterns, we can observe that the number of infected drops when initially infected quarantine, followed by a smooth transmission in the second phase when demand still fluctuates ($p=0.85$) or halted immediately ($p>0.95)$.
    Typically, stabilising the demand halts the epidemic faster. For p>0.85 epidemic is over in less than 50 days, while for p=0.65 it can remain active after 100 days (regardless number of initially infected).
    Despite a clear and strong trend, some simulated outbreaks do not follow the same patterns. We can observe for example an exceptionally high number of infections for p=0.8 starting from 2 infections when a highly connected hub got infected; quickly halted spreading from 10 infections at p = 0.65; or second wave at p=0.75 and 10 initial infections.}
    \label{fig:spreading}
\end{figure}

\begin{figure}
    \centering
    \includegraphics[width=0.7\textwidth]{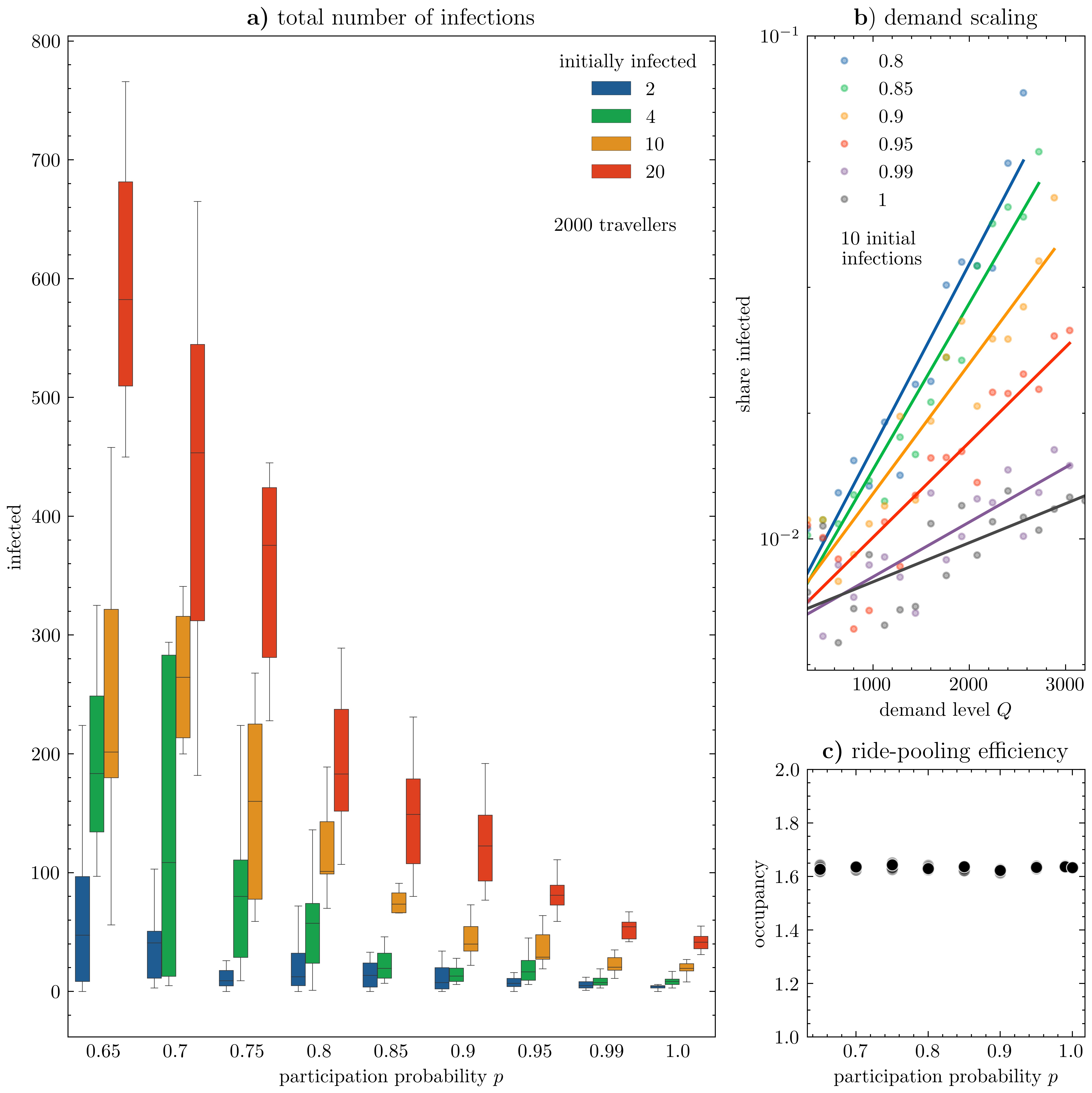}
    \caption{\small  \textbf{a)} Number of eventually infected travellers for varying demand stability $p$ and initially infected travellers. Distributions based on 20 replications (mean within interquantile box and whiskers from min to max values).
    Initially infected 20 travellers may spread the virus to almost 40\% of the population (800 out of 2000 travellers). Yet as long as demand becomes stabilised, outbreaks start being contained. Even a large number of initially infected does not reach more than 10\% of the population if demand stability is set to 90\% and is eventually contained below 60 travellers (3\%) for fully stable demand.
    The variability of outbreaks also decreases as the demand stabilises: 10 initial infections may reach between 40 and 100 travellers if $p=0.9$, while the range expands from 50 to over 400 for $p=0.7$. The lower bound increases when the number of initially infected is high, making outbreaks more predictable, unlike the ones starting from a small number of infections, for which variability is greater. Importantly, stabilizing the demand does not reduce the efficiency, as we report in \textbf{(c)}, where the mean occupancy (key efficiency indicator of ride-pooling) remains stable as demand stabilizes. 
    Notably, the importance of demand stabilization increases with the demand level as we demonstrate on panel \textbf{(b)} which shows share of infected travellers changing with a demand for various $p$'s. Each dot is the average from 20 replications. For all the values of $p$ share of infected individuals scales with the demand level as $A\exp(\alpha Q)$, marked as trendlines on \textbf{(b)}, with p-values and $R^2$ reported in the text.  For demand levels below 1500, the virus rarely reaches more than 4\% of the travellers. In contrast, when the demand level is 2500, the epidemic may reach up to 10\% when $p=0.8$ or be contained below 2\% for a stable demand ($p>0.9$), which underlines the importance of control measures for higher demand levels.}
    \label{fig:boxplots}
\end{figure}

\begin{figure}
    \centering
    \includegraphics[width=\textwidth]{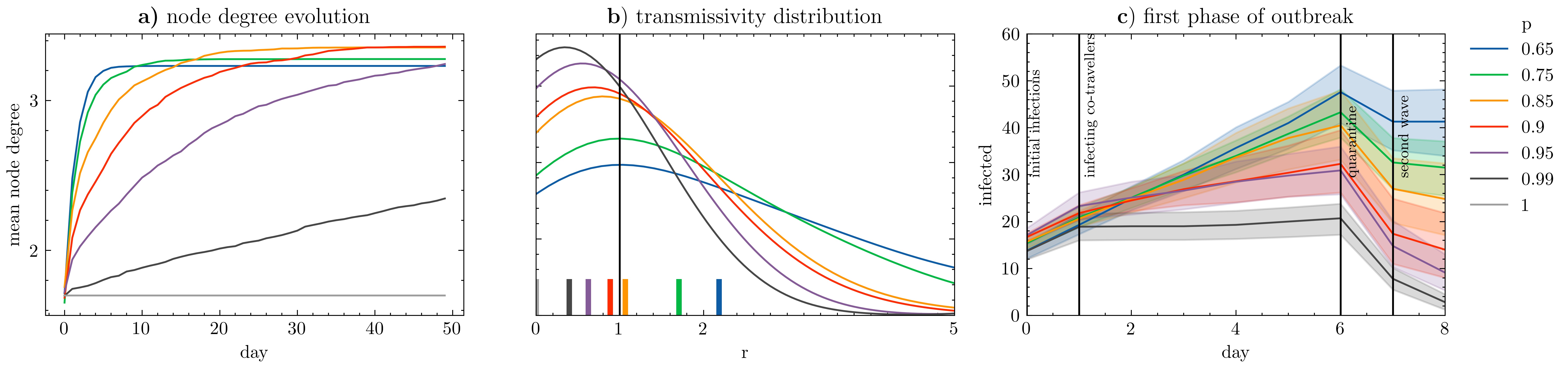}
    \caption{\small a) Average node degree in the evolving contact networks. Regardless demand stability $p$, an average traveller is linked to 1.7 other travellers each day. Yet if the demand is unstable it evolves, after 10 days it reaches 1.9 if $p=0.99$, 2.5 if $p=0.95$ and goes beyond 3 if $p<0.8$.
    b) Mean transmission rate $r$ (number of new infections per infected) distributions. The long tails for low demand stability reveal the super-spreaders (transmitting to 5 and more travellers). For a stable demand initial infections does not manage to transmit a disease effectively, eventually reducing transmissivity below 1 when $p>0.9$. 
    c) Insights into the first phase of the epidemic outbreak in the case of 10 initial infections. When first infected travellers are diagnosed after 7 days, their accumulated contact network may vary from 18 to over 60 infected travellers. If contact tracing and mitigation strategies are put in place, already infected travellers may be identified and quarantined before the second outbreak after day 7}
    \label{fig:3plots}
\end{figure}

\begin{figure}
    \centering
    \includegraphics[width=0.99\textwidth]{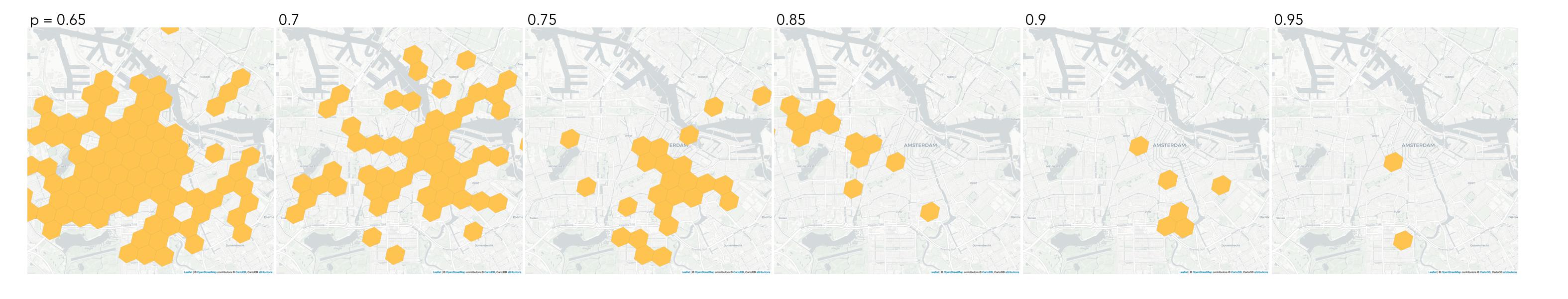}
    \caption{\small An illustration of the spatial extent of epidemic outbreaks originating from two initial infections. A major part of Amsterdam becomes infected for spontaneous demand (left), while it remains spatially contained as the demand stabilises (right). For stable demand ($p>0.8$) the geographical boundaries are confined, while otherwise, the virus crosses the river Ij and reaches also the north parts of Amsterdam.}
    \label{fig:spatial}
\end{figure}

We present the results through epidemic evolution plots (Fig. \ref{fig:spreading}) for various settings of demand stability and number of initial infections, summarised with boxplots of the total number of infections in Fig. \ref{fig:boxplots}a. On fig \ref{fig:boxplots}c we trace the efficiency of ride-pooling across scenarios. In Fig. \ref{fig:boxplots}b we explore spreading for increasing demand levels and reveal exponential growth of the share of infected characterized by fitting coefficients scaling linearly with participation probability $p$ (Fig. \ref{fig:boxplots}c).  To understand the impact of demand stability on the course of outbreaks we plot node degree evolution in Fig. \ref{fig:3plots}a and the distribution of transmissivity in Fig. \ref{fig:3plots}b, further visualised in terms of its spatial distributions in Fig. \ref{fig:spatial}. To demonstrate the potential to control the outbreak, we display its first phase in Fig. \ref{fig:3plots}c.

\paragraph{Outbreaks} As long as the demand is unstable and varies considerably from one day to the other, the virus may outbreak even when only a small number of passengers are initially infected. Outbreaks starting from two infections are highly variable (Fig. \ref{fig:spreading} lower left).
For initial spreaders located centrally in a highly connected giant component of the contact network (Fig. \ref{fig:graf}) the virus outbreaks and eventually infects over 250 travellers during the course of the spreading, whereas outbreaks from disconnected part of network can be naturally contained and halted already after 7 infections (Fig. \ref{fig:boxplots}a). An outbreak starting from 20 initial spreaders is always devastating and reaches from 450 up to almost 800 travellers, it needs only few days to reach 100 cases (Fig. \ref{fig:spreading} upper left). 
Such fast and prevalent spreading can be attributed to a gradually evolving contact network, where each additional day may bring opportunities to be pooled with a new set of travellers, extending the accumulated contact network (Fig. \ref{fig:3plots}a). Consequently, despite having a low mean node degree (i.e. sharing with few travellers at the time), some travellers become hubs, spreading the virus to over 10 travellers, resulting with a long tail of the transmission distribution (fig. \ref{fig:3plots}b). With low demand stability even two infections may spatially penetrate to all parts of the Amsterdam area, whereas for stable demand the virus may be contained spatially and not spread beyond its original community (fig. \ref{fig:spatial}).

\paragraph{Scaling for the demand level}
We experiment with changing the demand levels, gradually increasing it from from 100 to 3000 travellers (\ref{fig:boxplots}b). For low demand levels the virus cannot spread since the potential shareability network remains disconnected, i.e. with no giant component (few matches between sparse travellers are found and trips are rarely pooled). Thus, below the critical mass of ride pooling, stabilizing demand has a limited impact. However, as soon as the shareability network includes a larger number of connections (thanks to more compatible trip groups in the demand set) spreading is triggered and the importance of controlling becomes evident.  

The relation between the share of infected individuals $n_i$ and the demand level $Q$ can be fitted with an exponential function $n_i = A \exp(\alpha Q)$, allowing to make predictions about the number of people reached by the virus for higher values of $Q$ than those explored in our study.
For all values of $p$ shown in Fig. \ref{fig:boxplots}b we get high statistical significance of coefficients $A$ and $\alpha$ (in all cases p-value $< .001$), with the coefficient of determination $R^2$ of $0.92, 0.93, 0.87, 0.81, 0.73$ and $0.52$ for demand stabilities $p$ of $0.8, 0.85, 0.9, 0.95 , 0.99$ and $1$ respectively. For less stable demand spreading is ubiquitous and thus less variable, while for stable demand spreading can  still remain contained, leading to significant variability in the results and lower goodness-of-fit.

\paragraph{Ride-pooling efficiency}
Ride-pooling needs a critical mass of demand to become efficient and sustainable. We report ride-pooling efficiency by means of the average occupancy $o$, i.e. ratio of passenger-kilometer hours to vehicle kilometer hours. In line with previous studies\cite{kucharski2020exact}, we find that occupancy is a function of demand levels, yet, notably, stabilizing the demand with our control parameter $p$ does not affect it. As long as the same number of travellers participates in pooling everyday, the efficiency remains more or less stable, as we report in fig. \ref{fig:boxplots}c, where 2000 travellers participating daily in the system yield the same occupancy regardless of the replication (dots) and demand stability (x-axis).

\paragraph{Control and mitigation} Results show that the virus may easily spread through the ride-pooling networks infecting the majority of the population with a low epidemic threshold (20 initial spreaders may infect up to 800 out of 2000 travellers). While the ride-pooling service provider cannot control for the initial share of infected traveller, nor the incubation and recovery periods, the ride-pooling demand may be controlled to mitigate the virus spreading. Specifically, we show that imposing  fixed matches by means of a more stable demand level - solely by controlling for $p$, without making amendments to the matching algorithm itself - can mitigate the spreading and bring it to halt (Figs. \ref{fig:spreading} \ref{fig:boxplots}, \ref{fig:3plots}, \ref{fig:spatial}). We demonstrate that matching and its stability is key to halt epidemics in ride-pooling networks. It can be used proactively in the the design of the real-time matching algorithm. 

Moreover, if contact tracing apps are used, when a traveller is diagnosed not only s/he has to quarantine, but also his/her traced contact network over the relevant period of time can be identified and eventually isolated. As we show in Fig. \ref{fig:3plots}c, 10 initial infections will spread to a maximum of 60 travellers prior to diagnosis, which seems to be feasible to trace back, isolate and halt spreading.

\paragraph{Study limitations and caveats}
We aim at revealing the universal patterns characterising the spreading of a virus in ride-pooling networks, yet our findings shall be considered with caveats. Namely, we simulate only a subset of daily mobility patterns (afternoon commute), from many sources of non-recurrence present in travel patterns we picked-up one (participation probability), which we found sufficient to reproduce its impact, while role of others (varying and fluctuating travel modes, destinations, departure times etc.) may be similar or potentially even stronger. While the shareability network in the morning commute is likely to be similar (inverse of afternoon), other, non-commuting trips, will likely yield a different shareability network, catalysing the spreading to the new co-travellers. Nonetheless, our results are valid for systems with regular users with symmetric demand patterns in the morning and afternoon. 
Moreover, drivers are assumed not to be spreaders (which seems plausible in the context of ad-hoc made shields isolating many of ride-sourcing drivers from travellers). We applied a fixed and deterministic epidemiological model in terms of the infection probability, incubation period and quarantining, since reliable estimates of those parameters distribution are not reported yet. 
Despite, we claim that the main message holds true for general urban networks: without intervention ride-pooling significantly contributes to virus spreading, while fixing matches between co-travellers dramatically reduces transmission.

\section*{Discussion and conclusions}
Sharing a single vehicle with co-travellers during pandemics induces a risk to become exposed to viruses. Sadly, the risk extends beyond the fellow travellers one shares the ride within a single vehicle, mainly due to the accumulated contact graph resulting from day to day variations. Regardless of the number of initial infections, the upper bound of outbreaks in spontaneous ride-pooling networks is high. Even two initial infections may lead to hundreds of cases across the network. We did not observe spatial nor topological limits to the spreading and disease starting from only two initial infections managed to reach most parts of the Amsterdam's area. 

In plausible demand and behavioural settings, if a generic ride-pooling system reaches a critical mass, the travellers become densely connected through the shareability network so that the virus transmits easily through the giant component without clearly visible epidemic thresholds. Only travellers belonging to isolated communities are left unaffected. The pace at which it spreads, however, is low, requiring a long time until virus penetrates across the network. Nonetheless, the daily contact network with its low node degree and hub-free, will evolve due to spontaneity in the demand patterns. If each day, a slightly different pool of travellers decides to travel, this will yield a new matching and resulting with a new shareability, contributing to the accumulated number of contacts steadily growing over time. The slow pace evolution of contact networks becomes beneficial when tracking measures are applied and we can trace past co-travellers for each diagnosed spreader. Even in a highly spontaneous ride-pooling networks, 10 initial infections manage to transmit to only 60 within the 7 day incubation period, which seems to be feasible to trace, isolate and halt the spreading, specifically given the app-based operations of the mobility platform, presumably storing travellers' traces anyhow. Otherwise, if not halted early, epidemics may evolve unhampered and randomly. Depending on the location of initial infections, the epidemic may die-out as well as outbreak, making it potentially risky and uncertain.

Notably, we can substantially limit spreading by sacrificing the spontaneity offered by the ride-pooling service. If we enforce the same matching and fix the pools of co-travellers, spreading is efficiently mitigated and even 20 initial infections remain manageable to be contained. With one, clearly controllable parameter we can reduce the outbreak of viruses. If translated to platform operations, this can become an efficient management and control measure, adjustable along with other country-wide pandemic measures. This may contribute to the provision of a safe shared-mobility alternative in the presence of public health fears and risks. Future research may modify the matching algorithm itself so as to favor the matching of travellers that have already travelled together in past rides. Such an approach is expected to allow for virus spreading reductions even when the demand pattern is subject to large day-to-day variations. Finally, if tracing is combined with fixed pooling, the system's safety may further improve, making ride-pooling a promising intermediate mobility solution for the pandemic world.

 Although the presented methodology has been illustrated on the case of Amsterdam ride-pooling it is essential to emphasise its general applicability to examine, in a non-invasive way, the likely outcomes of different underlying topologies on the way the virus spreads through the network. In this sense it opens space for discussion of potential alterations of practical ride-pooling systems on one side and theoretical studies on the other one. Although our study considers a limited number of rides, it is widely acknowledged that cities and their properties are connected with scaling laws \cite{Bettencourt2007} even if the form and the details of the methodology behind these relations is questioned \cite{Leitao2016,Altmann2020}. Typically such laws\cite{Sienkiewicz2005} may associate a certain index $x$ with city population size $M$ by an allometric scaling $x \sim M^{\alpha}$ . More importantly the existence of scaling laws has also been proven in the case of ride-sharing networks with respect to shareability \cite{Tachet2017}, visitation frequency \cite{Chen2018} or lately ride-sharing efficiency \cite{Molkenthin2020}. In view of this we may assume that our results regarding the number of infected individuals as a function of the demand level $Q$ should hold for larger systems emphasising the necessity to stabilize the demand.

\section*{Code and data availability}
The code to generate the shareability network from a given demand pattern and then to reproduce the epidemic simulations is available at the public GitHub repository (\url{github.com/rafalkucharskiPK/ExMAS}). The experimental results data is available under \url{doi.org/10.4121/14140616.v1}. The network data was obtained from Open Street Map with \texttt{osmnx}, Amsterdam travel demand was derived from Albatross data set \cite{arentze2004learning}.

\section*{Acknowledgements}
This research was supported by the CriticalMaaS project (grant number 804469), which is financed by the European Research Council and Amsterdam Institute for Advanced Metropolitan Solutions and the Transport Institute of TU Delft. This research was funded by National Science Centre in Poland program OPUS 19 (grant number 2020/37/B/HS4/01847) and by IDUB against COVID-19 project granted by Warsaw University of Technology under the program Excellence Initiative: Research University (IDUB).

\section*{Author contributions statement}
R.K. and O.C. conceived the experiment(s),  R.K. conducted the experiment(s), R.K O.C and J.S. analysed the results.  All authors reviewed the manuscript. 

\section*{Methods}

\paragraph{Travel demand data}
We run a series of experiments on a travel dataset available for Amsterdam from a nation-wide activity-based model \cite{arentze2004learning}, with a single trip defined as a combination of its origin $o_i$, destination $d_i$ and desired pick-up time $t^p_i$:
\begin{equation}
Q_i = (o_i,d_i,t_i).
\label{trip}
\end{equation}
The dataset contains over 240 thousand trips conducted within the boundaries of Amsterdam on a representative working day, which we filter to afternoon (2PM - 6PM) trips longer than one kilometer. We use 3200 passenger trip requests for the experiments, 2000 of which participates in the pooling on any given day. The pool of travellers from which we sample the daily demand is controlled using $p$, based on which each day we draw from the pool of 2000/$p$ travellers.

\paragraph{Ride-pooling algorithm}
To identify attractive pooled rides we use the \texttt{ExMAS}\cite{kucharski2020exact} algorithm (publicly available python library), which for a given network (\texttt{osmnx} graph), travel demand, behavioral parameters (like willingness-to-share) and system parameters (pooling discount) identifies all feasible shared rides and then constructs a shareability network (fig. \ref{fig:graf}a) to finally optimally match trips into shared rides (fig. \ref{fig:graf}b).

It generates the so-called shareability network, linking two kinds of nodes: travellers and rides. Traveller $i$ is linked to a feasible ride $r$ if and only if s/he finds it attractive, which we express as the probability that ride utility $U_{i,r}$ - reflecting the extent to which delays and detours $\delta_{r,i}$ imposed by sharing are compensated by a discounted ride fare $\lambda$ under traveller's behavioural parameters $\beta_i$ (value of time and willingness-to-share) - is greater than travellers' attractiveness threshold $\epsilon_i$. 
The theoretical number of shared-rides explodes combinatorically with the number of travellers (e.g. 2000 travellers can be matched into $4.65\times10^{20}$ theoretically feasible trips shared by up to five passengers). This can be made tractable by considering only attractive rides, which is governed on one hand by travellers preferences $\beta_i$ (i.e. individual trade-offs between longer ride and discounted price) and on the other hand by service design $\lambda$ (controlled through the discount offered by the platform for sharing) and $\epsilon_i$ expressing the quality of alternatives for ride-pooling (private ride-hailing, or public transport and/or bike), further detailed in\cite{kucharski2020exact}.
Importantly, the shareability network is composed of feasible rides only, expressed with $F_r$, being one if the ride is attractive for all travellers sharing it and zero otherwise. We formalize the shareability network with a link formation $l_{i,r}$ formula, combining ride feasibility and attractiveness as follows:
\begin{equation}
    l_{i,r} = F_r \cdot \Pr (U_{i,r} = U(i, \beta_i, \delta_{r,i}, \lambda) > \epsilon_i)  \label{bool}
\end{equation}

\paragraph{Matching travellers to attractive shared rides}
Each traveller may be linked to multiple rides and the resulting shareability network is typically highly connected, characterized by the formation of communities and hubs (Fig. \ref{fig:graf}a). While the shareability network denotes the potential to share a ride, on any given day travellers are matched to exactly one particular shared-ride (Fig. \ref{fig:graf}b).  

To address this, we formulate a binary traveller-ride assignment problem, where each traveller $i$ is unilaterally assigned to a ride $r$ and the assignment yields the minimal costs. It is formulated as a problem of determining a binary vector $x_r$, an assignment variable equal to one if a ride is selected and zero otherwise (eq. \ref{eq:constraint2}). The objective of this deterministic assignment are ride costs $c_r$, multiplied by the assignment variable $x_r$, aggregated for all rides (eq. \ref{Obj}). 
\par
Such an assignment satisfies the constraint of assigning each traveller to exactly one ride, obtained through the row-wise sum for assignment variable $x_r$ and traveller-ride incidence matrix $I_{i,r}$. The latter is a binary matrix, where each entry is one if ride $r$ serves traveller $i$ and zero otherwise (eq. \ref{eq:constraint1}).
Eventually, the solution to the problem (eq. \ref{Obj}) is the subset $\mathbf{R^*}$ of feasible rides $\mathbf{R}$ such that $x_r =1$ $\forall r \in \mathbf{R^*}$. 
We express the shareability problem as the following program:
\begin{linenomath*}
\begin{subequations}
\begin{alignat}{2}
&\!\min        &\qquad& \sum_{r\in \mathbf{R}} c_r x_r \label{Obj}\\
&\text{subject to} &      & \sum_{i \in \mathbf{Q}} I_{i,r} x_r = 1 , \forall i  \label{eq:constraint1}\\
&                  &      & x_r \in \lbrace 0,1 \rbrace .\label{eq:constraint2}
\end{alignat}
\end{subequations}
\end{linenomath*}

Although matching problem (eq. \ref{Obj}) can be read  as the set cover problem\cite{raz1997sub}, which is known to be NP-Hard, real-life ride-pooling situations usually yield configurations managed by standard solvers (like in \cite{kucharski2020exact,alonso2017demand}).

\paragraph{Contact network}
On any given day, the contact graph is composed of connecting each ride to all travellers that have shared (part of) it. 
Notably, the contact network evolves over time, primarily due to the different pool of travellers being matched on any given day. Hence, this representation allows simulating an epidemic outbreak by analyzing potential transmissions between travellers that have shared rides with other travellers over the course of the analysis period. In our model the contact network changes from day to day due to one or more of the following reasons: (i) infected travellers quarantine (which may catalyse spreading as quarantined travellers are replaced by susceptible ones, who will get infected) (ii) recovered travellers return to the system (which impedes spreading as recovered travellers restore to previous, optimal matches, already penetrated by the virus) or (iii) daily variations in travel demand as travellers decide not to use ride-hailing on a given day (for example because they opt for an alternative mode). We represent the daily participation, central endogenous variable of the model, through the \textit{demand stability} parameter $p$ in our experiments. Each day we update the pool of travellers (using the daily participation formula $F^d_i = \Pr(p) \cdot (1 -K^d_i)$ which combines the participation probability $p$ and quarantined travellers on day $K_d^i$). This, in turn, results with updating the pool of rides feasible on a given day (composed only of travellers present in the daily pool). The contact network will then evolve as travellers are matched to new rides when their co-travellers are quarantined or absent.

\paragraph{Epidemic model}
We adopt a \textbf{SIQR} model to represent the four compartments characteristic of the COVID-19 pandemic: Susceptible ($S$), Infected ($I$), Quarantined ($Q$) and Recovered ($R$), recently, directly applied to tackle COVID-19 propagation in other studies (Italy\cite{Pedersen2020} and Japan\cite{Odagaki2020}). Following the argumentation of Pedersen and Meneghini\cite{Pedersen2020} we do not explicitly designate the E state, given the evidence suggesting that the COVID-19 virus can be propagated without first exhibiting visual symptoms. The SIQR model was first introduced by Feng and Thieme in 1995\cite{Feng1995} and then examined in detail by Hethcote {\it et al.}\cite{Hethcote2002}. Previous studies focused on mathematical aspects of the model (e.g., oscillations\cite{Feng1995}, stability analysis\cite{Hethcote2002} or the role of stochastic noise\cite{Cao2019}). While the aggregate epidemiological properties of the SIQR model are well studied, studies taking into account the underlying network structure and its evolution are scarce. 

The phenomenon central to this paper is driven by the structure and evolution of the contact network, rather than by the parameters of the epidemic model. We, therefore, adopt a deterministic model where infected travellers infect all of their co-travellers with a probability of 1. For the sake of clarity, unlike SEIR models, we assume that all exposed inevitably become infected, all of which quarantine and recover after certain incubation and recovery periods (we use here the latest reliable findings suggesting, respectively, 7 and 14 for COVID-19\cite{lauer2020incubation}).
This ubiquitous spread over the contact network may be seen as a pessimistic upper-bound of the spreading process, yet in the view of recent pandemics\cite{liu2020secondary}, sharing a vehicle with infected co-traveller is expected to yield a high contagion risk. Furthermore, the focus of this study is on spreading across the network and over multiple vehicles and rides rather than within vehicle transmission probabilities. Future medical estimates of the latter can be embedded into the analysis performed in this study as soon as those are made available to refine our model specifications and thus obtain more precise estimates. Our findings should therefore be considered an upper-bound of the epidemiological consequences of virus spreading in ride-pooling systems.

\paragraph{Modelling framework}

The ExMAS ride-pooling algorithm is embedded within the day-to-day loop characteristic to epidemiological model. The simulation initializes with a trip demand set composed of all the travellers that may consider ride-pooling on any given day during the course of the simulated epidemic outbreak, to further allow embedding the participation probability $p$. Before entering the main epidemic loop, we identify all feasible pooled rides (fig. \ref{fig:graf}a) - to determine potential co-travellers that any given traveller may encounter during the course of an epidemic outbreak. We create the complete shareability graph by applying equation (\ref{bool}) with $\epsilon$ corresponding to a private, non-shared ride alternative in a deterministic model.

Following this initialisation phase, we then enter the main simulation loop. We start with assigning initial infectors - drawn in random by sampling a pre-defined number of initially infected, which is treated as random input and vary from one replication to another. Next, we enter the day-to-day simulation: every day we first determine the daily ride-pooling demand. We assume that only a subset of travellers actually participates in the ride-pooling system on any given day, i.e. every day we sample a given number of travellers from the total latent demand. We fix the demand to 2000 everyday in our experiments to ease comparisons (except fig. \ref{fig:boxplots}b where we experiment with various demand levels). Those travellers are then matched to identify the realization of shared-rides on a given day.
Everyday we apply the SIQR model with transitions taking place when:

(a) infected travellers infect their susceptible co-riders ($S\to I$),

(b) infected travellers are quarantined after the incubation period ($I\to Q$), 

(c) travellers recover after the quarantine and acquire complete immunity to the virus ($Q\to R$). 

For any given day, the model outputs information about the number of travellers in each state (S-I-Q-R) and newly infected travellers, based on which we can reproduce epidemic spreading profiles. The loop terminates when all the infected travellers are quarantined (there are no active infections).

\bibliography{references}

\begin{thebibliography}{10}
\urlstyle{rm}
\expandafter\ifx\csname url\endcsname\relax
  \def\url#1{\texttt{#1}}\fi
\expandafter\ifx\csname urlprefix\endcsname\relax\def\urlprefix{URL }\fi
\expandafter\ifx\csname doiprefix\endcsname\relax\def\doiprefix{DOI: }\fi
\providecommand{\bibinfo}[2]{#2}
\providecommand{\eprint}[2][]{\url{#2}}

\bibitem{Acuto2020}
\bibinfo{author}{Acuto, M.} \emph{et~al.}
\newblock \bibinfo{journal}{\bibinfo{title}{{Seeing COVID-19 through an urban
  lens}}}.
\newblock {\emph{\JournalTitle{Nature Sustainability}}}
  \doiprefix\url{10.1038/s41893-020-00620-3} (\bibinfo{year}{2020}).

\bibitem{He2020}
\bibinfo{author}{He, G.}, \bibinfo{author}{Pan, Y.} \& \bibinfo{author}{Tanaka,
  T.}
\newblock \bibinfo{journal}{\bibinfo{title}{{The short-term impacts of COVID-19
  lockdown on urban air pollution in China}}}.
\newblock {\emph{\JournalTitle{Nature Sustainability}}}
  \doiprefix\url{10.1038/s41893-020-0581-y} (\bibinfo{year}{2020}).

\bibitem{muller2020mobility}
\bibinfo{author}{Muller, S.~A.}, \bibinfo{author}{Balmer, M.},
  \bibinfo{author}{Neumann, A.} \& \bibinfo{author}{Nagel, K.}
\newblock \bibinfo{journal}{\bibinfo{title}{Mobility traces and spreading of
  covid-19}}.
\newblock {\emph{\JournalTitle{medRxiv}}}  (\bibinfo{year}{2020}).

\bibitem{UITP}
\bibinfo{author}{UITP}.
\newblock \bibinfo{title}{Management of covid-19 guidelines for public
  transport operators} (\bibinfo{year}{2020}).

\bibitem{gkiotsalitis2020optimal}
\bibinfo{author}{Gkiotsalitis, K.} \& \bibinfo{author}{Cats, O.}
\newblock \bibinfo{journal}{\bibinfo{title}{Optimal frequency setting of metro
  services in the age of covid-19 distancing measures}}.
\newblock {\emph{\JournalTitle{arXiv preprint arXiv:2006.05688}}}
  (\bibinfo{year}{2020}).

\bibitem{tirachini2020covid}
\bibinfo{author}{Tirachini, A.} \& \bibinfo{author}{Cats, O.}
\newblock \bibinfo{journal}{\bibinfo{title}{Covid-19 and public transportation:
  Current assessment, prospects, and research needs}}.
\newblock {\emph{\JournalTitle{Journal of Public Transportation}}}
  \textbf{\bibinfo{volume}{22}}, \bibinfo{pages}{1} (\bibinfo{year}{2020}).

\bibitem{guerriero2020health}
\bibinfo{author}{Guerriero, C.}, \bibinfo{author}{Haines, A.} \&
  \bibinfo{author}{Pagano, M.}
\newblock \bibinfo{journal}{\bibinfo{title}{Health and sustainability in
  post-pandemic economic policies}}.
\newblock {\emph{\JournalTitle{Nature Sustainability}}} \bibinfo{pages}{1--3}
  (\bibinfo{year}{2020}).

\bibitem{alonso2017demand}
\bibinfo{author}{Alonso-Mora, J.}, \bibinfo{author}{Samaranayake, S.},
  \bibinfo{author}{Wallar, A.}, \bibinfo{author}{Frazzoli, E.} \&
  \bibinfo{author}{Rus, D.}
\newblock \bibinfo{journal}{\bibinfo{title}{On-demand high-capacity
  ride-sharing via dynamic trip-vehicle assignment}}.
\newblock {\emph{\JournalTitle{Proceedings of the National Academy of
  Sciences}}} \textbf{\bibinfo{volume}{114}}, \bibinfo{pages}{462--467}
  (\bibinfo{year}{2017}).

\bibitem{kucharski2020exact}
\bibinfo{author}{Kucharski, R.} \& \bibinfo{author}{Cats, O.}
\newblock \bibinfo{journal}{\bibinfo{title}{Exact matching of attractive shared
  rides (exmas) for system-wide strategic evaluations}}.
\newblock {\emph{\JournalTitle{Transportation Research Part B:
  Methodological}}} \textbf{\bibinfo{volume}{139}}, \bibinfo{pages}{285 --
  310}, \doiprefix\url{https://doi.org/10.1016/j.trb.2020.06.006}
  (\bibinfo{year}{2020}).

\bibitem{Yang2020}
\bibinfo{author}{Yang, B.} \emph{et~al.}
\newblock \bibinfo{journal}{\bibinfo{title}{{Phase transition in taxi dynamics
  and impact of ridesharing}}}.
\newblock {\emph{\JournalTitle{Transportation Science}}}
  \textbf{\bibinfo{volume}{54}}, \bibinfo{pages}{250--273},
  \doiprefix\url{10.1287/trsc.2019.0915} (\bibinfo{year}{2020}).
\newblock \eprint{1801.00462}.

\bibitem{Riascos2020}
\bibinfo{author}{Riascos, A.~P.} \& \bibinfo{author}{Mateos, J.~L.}
\newblock \bibinfo{journal}{\bibinfo{title}{{Networks and long-range mobility
  in cities: A study of more than one billion taxi trips in New York City}}}.
\newblock {\emph{\JournalTitle{Scientific Reports}}}
  \textbf{\bibinfo{volume}{10}}, \bibinfo{pages}{1--14},
  \doiprefix\url{10.1038/s41598-020-60875-w} (\bibinfo{year}{2020}).

\bibitem{Tachet2017}
\bibinfo{author}{Tachet, R.} \emph{et~al.}
\newblock \bibinfo{journal}{\bibinfo{title}{{Scaling Law of Urban Ride
  Sharing}}}.
\newblock {\emph{\JournalTitle{Scientific Reports}}}
  \textbf{\bibinfo{volume}{7}}, \bibinfo{pages}{42868},
  \doiprefix\url{10.1038/srep42868} (\bibinfo{year}{2017}).

\bibitem{Chen2018}
\bibinfo{author}{Chen, X.~M.}, \bibinfo{author}{Chen, C.}, \bibinfo{author}{Ni,
  L.} \& \bibinfo{author}{Li, L.}
\newblock \bibinfo{journal}{\bibinfo{title}{{Spatial visitation prediction of
  on-demand ride services using the scaling law}}}.
\newblock {\emph{\JournalTitle{Physica A: Statistical Mechanics and its
  Applications}}} \textbf{\bibinfo{volume}{508}}, \bibinfo{pages}{84--94},
  \doiprefix\url{https://doi.org/10.1016/j.physa.2018.05.005}
  (\bibinfo{year}{2018}).

\bibitem{santi2014quantifying}
\bibinfo{author}{Santi, P.} \emph{et~al.}
\newblock \bibinfo{journal}{\bibinfo{title}{Quantifying the benefits of vehicle
  pooling with shareability networks}}.
\newblock {\emph{\JournalTitle{Proceedings of the National Academy of
  Sciences}}} \textbf{\bibinfo{volume}{111}}, \bibinfo{pages}{13290--13294}
  (\bibinfo{year}{2014}).

\bibitem{liu2020secondary}
\bibinfo{author}{Liu, Y.}, \bibinfo{author}{Eggo, R.~M.} \&
  \bibinfo{author}{Kucharski, A.~J.}
\newblock \bibinfo{journal}{\bibinfo{title}{Secondary attack rate and
  superspreading events for sars-cov-2}}.
\newblock {\emph{\JournalTitle{The Lancet}}} \textbf{\bibinfo{volume}{395}},
  \bibinfo{pages}{e47} (\bibinfo{year}{2020}).

\bibitem{DellaRossa2020}
\bibinfo{author}{{Della Rossa}, F.} \emph{et~al.}
\newblock \bibinfo{journal}{\bibinfo{title}{{A network model of Italy shows
  that intermittent regional strategies can alleviate the COVID-19 epidemic}}}.
\newblock {\emph{\JournalTitle{Nature Communications}}}
  \textbf{\bibinfo{volume}{11}}, \bibinfo{pages}{5106},
  \doiprefix\url{10.1038/s41467-020-18827-5} (\bibinfo{year}{2020}).

\bibitem{LiuZhu2020}
\bibinfo{author}{Liu, Z.} \emph{et~al.}
\newblock \bibinfo{journal}{\bibinfo{title}{{Near-real-time monitoring of
  global CO2 emissions reveals the effects of the COVID-19 pandemic}}}.
\newblock {\emph{\JournalTitle{Nature Communications}}}
  \textbf{\bibinfo{volume}{11}}, \bibinfo{pages}{5172},
  \doiprefix\url{10.1038/s41467-020-18922-7} (\bibinfo{year}{2020}).

\bibitem{editorial}
\bibinfo{journal}{\bibinfo{title}{Developing infectious disease surveillance
  systems}}.
\newblock {\emph{\JournalTitle{Nature Communications}}}
  \textbf{\bibinfo{volume}{11}}, \bibinfo{pages}{4962},
  \doiprefix\url{10.1038/s41467-020-18798-7} (\bibinfo{year}{2020}).

\bibitem{pastor2015epidemic}
\bibinfo{author}{Pastor-Satorras, R.}, \bibinfo{author}{Castellano, C.},
  \bibinfo{author}{Van~Mieghem, P.} \& \bibinfo{author}{Vespignani, A.}
\newblock \bibinfo{journal}{\bibinfo{title}{Epidemic processes in complex
  networks}}.
\newblock {\emph{\JournalTitle{Reviews of modern physics}}}
  \textbf{\bibinfo{volume}{87}}, \bibinfo{pages}{925} (\bibinfo{year}{2015}).

\bibitem{wang2016}
\bibinfo{author}{Wang, Z.} \emph{et~al.}
\newblock \bibinfo{journal}{\bibinfo{title}{Statistical physics of
  vaccination}}.
\newblock {\emph{\JournalTitle{Physics Reports}}}
  \textbf{\bibinfo{volume}{664}}, \bibinfo{pages}{1 -- 113},
  \doiprefix\url{https://doi.org/10.1016/j.physrep.2016.10.006}
  (\bibinfo{year}{2016}).

\bibitem{Sienkiewicz2005}
\bibinfo{author}{Sienkiewicz, J.} \& \bibinfo{author}{Ho{\l}yst, J.~A.}
\newblock \bibinfo{journal}{\bibinfo{title}{Statistical analysis of 22 public
  transport networks in poland}}.
\newblock {\emph{\JournalTitle{Phys. Rev. E}}} \textbf{\bibinfo{volume}{72}},
  \bibinfo{pages}{046127}, \doiprefix\url{10.1103/PhysRevE.72.046127}
  (\bibinfo{year}{2005}).

\bibitem{Barthelemy2011}
\bibinfo{author}{Barth{\'{e}}lemy, M.}
\newblock \bibinfo{journal}{\bibinfo{title}{{Spatial networks}}}.
\newblock {\emph{\JournalTitle{Physics Reports}}}
  \textbf{\bibinfo{volume}{499}}, \bibinfo{pages}{1--101},
  \doiprefix\url{https://doi.org/10.1016/j.physrep.2010.11.002}
  (\bibinfo{year}{2011}).

\bibitem{Gallotti2015}
\bibinfo{author}{Gallotti, R.} \& \bibinfo{author}{Barthelemy, M.}
\newblock \bibinfo{journal}{\bibinfo{title}{{The multilayer temporal network of
  public transport in Great Britain}}}.
\newblock {\emph{\JournalTitle{Scientific Data}}} \textbf{\bibinfo{volume}{2}},
  \bibinfo{pages}{140056}, \doiprefix\url{10.1038/sdata.2014.56}
  (\bibinfo{year}{2015}).

\bibitem{Aleta2020}
\bibinfo{author}{Aleta, A.}, \bibinfo{author}{Hu, Q.}, \bibinfo{author}{Ye,
  J.}, \bibinfo{author}{Ji, P.} \& \bibinfo{author}{Moreno, Y.}
\newblock \bibinfo{journal}{\bibinfo{title}{A data-driven assessment of early
  travel restrictions related to the spreading of the novel covid-19 within
  mainland china}}.
\newblock {\emph{\JournalTitle{Chaos, Solitons \& Fractals}}}
  \textbf{\bibinfo{volume}{139}}, \bibinfo{pages}{110068},
  \doiprefix\url{https://doi.org/10.1016/j.chaos.2020.110068}
  (\bibinfo{year}{2020}).

\bibitem{Chinazzi2020}
\bibinfo{author}{Chinazzi, M.} \emph{et~al.}
\newblock \bibinfo{journal}{\bibinfo{title}{The effect of travel restrictions
  on the spread of the 2019 novel coronavirus (covid-19) outbreak}}.
\newblock {\emph{\JournalTitle{Science}}} \textbf{\bibinfo{volume}{368}},
  \bibinfo{pages}{395--400}, \doiprefix\url{10.1126/science.aba9757}
  (\bibinfo{year}{2020}).
\newblock
  \eprint{https://science.sciencemag.org/content/368/6489/395.full.pdf}.

\bibitem{Kraemer2020}
\bibinfo{author}{Kraemer, M. U.~G.} \emph{et~al.}
\newblock \bibinfo{journal}{\bibinfo{title}{The effect of human mobility and
  control measures on the covid-19 epidemic in china}}.
\newblock {\emph{\JournalTitle{Science}}} \textbf{\bibinfo{volume}{368}},
  \bibinfo{pages}{493--497}, \doiprefix\url{10.1126/science.abb4218}
  (\bibinfo{year}{2020}).
\newblock
  \eprint{https://science.sciencemag.org/content/368/6490/493.full.pdf}.

\bibitem{Azizi2020}
\bibinfo{author}{Azizi, A.}, \bibinfo{author}{Montalvo, C.},
  \bibinfo{author}{Espinoza, B.}, \bibinfo{author}{Kang, Y.} \&
  \bibinfo{author}{Castillo-Chavez, C.}
\newblock \bibinfo{journal}{\bibinfo{title}{Epidemics on networks: Reducing
  disease transmission using health emergency declarations and peer
  communication}}.
\newblock {\emph{\JournalTitle{Infectious Disease Modelling}}}
  \textbf{\bibinfo{volume}{5}}, \bibinfo{pages}{12 -- 22},
  \doiprefix\url{https://doi.org/10.1016/j.idm.2019.11.002}
  (\bibinfo{year}{2020}).

\bibitem{xue2020}
\bibinfo{author}{Xue, L.} \emph{et~al.}
\newblock \bibinfo{journal}{\bibinfo{title}{A data-driven network model for the
  emerging covid-19 epidemics in wuhan, toronto and italy}}.
\newblock {\emph{\JournalTitle{Mathematical Biosciences}}}
  \textbf{\bibinfo{volume}{326}}, \bibinfo{pages}{108391},
  \doiprefix\url{https://doi.org/10.1016/j.mbs.2020.108391}
  (\bibinfo{year}{2020}).

\bibitem{Ciaperoni2020}
\bibinfo{author}{Ciaperoni, M.} \emph{et~al.}
\newblock \bibinfo{journal}{\bibinfo{title}{{Relevance of temporal cores for
  epidemic spread in temporal networks}}}.
\newblock {\emph{\JournalTitle{Scientific Reports}}}
  \textbf{\bibinfo{volume}{10}}, \bibinfo{pages}{12529},
  \doiprefix\url{10.1038/s41598-020-69464-3} (\bibinfo{year}{2020}).

\bibitem{Cacciapaglia2020}
\bibinfo{author}{Cacciapaglia, G.}, \bibinfo{author}{Cot, C.} \&
  \bibinfo{author}{Sannino, F.}
\newblock \bibinfo{journal}{\bibinfo{title}{{Second wave COVID-19 pandemics in
  Europe: a temporal playbook}}}.
\newblock {\emph{\JournalTitle{Scientific Reports}}}
  \textbf{\bibinfo{volume}{10}}, \bibinfo{pages}{15514},
  \doiprefix\url{10.1038/s41598-020-72611-5} (\bibinfo{year}{2020}).

\bibitem{volz2009epidemic}
\bibinfo{author}{Volz, E.} \& \bibinfo{author}{Meyers, L.~A.}
\newblock \bibinfo{journal}{\bibinfo{title}{Epidemic thresholds in dynamic
  contact networks}}.
\newblock {\emph{\JournalTitle{Journal of the Royal Society Interface}}}
  \textbf{\bibinfo{volume}{6}}, \bibinfo{pages}{233--241}
  (\bibinfo{year}{2009}).

\bibitem{Feng1995}
\bibinfo{author}{Feng, Z.} \& \bibinfo{author}{Thieme, H.~R.}
\newblock \bibinfo{journal}{\bibinfo{title}{Recurrent outbreaks of childhood
  diseases revisited: The impact of isolation}}.
\newblock {\emph{\JournalTitle{Mathematical Biosciences}}}
  \textbf{\bibinfo{volume}{128}}, \bibinfo{pages}{93 -- 130},
  \doiprefix\url{https://doi.org/10.1016/0025-5564(94)00069-C}
  (\bibinfo{year}{1995}).

\bibitem{Molkenthin2020}
\bibinfo{author}{Molkenthin, N.}, \bibinfo{author}{Schr{\"{o}}der, M.} \&
  \bibinfo{author}{Timme, M.}
\newblock \bibinfo{journal}{\bibinfo{title}{{Scaling Laws of Collective
  Ride-Sharing Dynamics}}}.
\newblock {\emph{\JournalTitle{Physical Review Letters}}}
  \textbf{\bibinfo{volume}{125}}, \bibinfo{pages}{248302},
  \doiprefix\url{10.1103/PhysRevLett.125.248302} (\bibinfo{year}{2020}).

\bibitem{aslak2019netwulf}
\bibinfo{author}{Aslak, U.} \& \bibinfo{author}{Maier, B.~F.}
\newblock \bibinfo{journal}{\bibinfo{title}{Netwulf: Interactive visualization
  of networks in python}}.
\newblock {\emph{\JournalTitle{Journal of Open Source Software}}}
  \textbf{\bibinfo{volume}{4}}, \bibinfo{pages}{1425} (\bibinfo{year}{2019}).

\bibitem{arentze2004learning}
\bibinfo{author}{Arentze, T.~A.} \& \bibinfo{author}{Timmermans, H.~J.}
\newblock \bibinfo{journal}{\bibinfo{title}{A learning-based transportation
  oriented simulation system}}.
\newblock {\emph{\JournalTitle{Transportation Research Part B:
  Methodological}}} \textbf{\bibinfo{volume}{38}}, \bibinfo{pages}{613 -- 633},
  \doiprefix\url{https://doi.org/10.1016/j.trb.2002.10.001}
  (\bibinfo{year}{2004}).

\bibitem{alonso2020determinants}
\bibinfo{author}{Alonso-Gonz{\'a}lez, M.~J.} \emph{et~al.}
\newblock \bibinfo{journal}{\bibinfo{title}{What are the determinants of the
  willingness to share rides in pooled on-demand services?}}
\newblock {\emph{\JournalTitle{Transportation}}} \bibinfo{pages}{1--33}
  (\bibinfo{year}{2020}).

\bibitem{alonso2020value}
\bibinfo{author}{Alonso-Gonz{\'a}lez, M.~J.}, \bibinfo{author}{van Oort, N.},
  \bibinfo{author}{Cats, O.}, \bibinfo{author}{Hoogendoorn-Lanser, S.} \&
  \bibinfo{author}{Hoogendoorn, S.}
\newblock \bibinfo{journal}{\bibinfo{title}{Value of time and reliability for
  urban pooled on-demand services}}.
\newblock {\emph{\JournalTitle{Transportation Research Part C: Emerging
  Technologies}}} \textbf{\bibinfo{volume}{115}}, \bibinfo{pages}{102621}
  (\bibinfo{year}{2020}).

\bibitem{LI2019330}
\bibinfo{author}{Li, W.}, \bibinfo{author}{Pu, Z.}, \bibinfo{author}{Li, Y.} \&
  \bibinfo{author}{{(Jeff) Ban}, X.}
\newblock \bibinfo{journal}{\bibinfo{title}{Characterization of ridesplitting
  based on observed data: A case study of chengdu, china}}.
\newblock {\emph{\JournalTitle{Transportation Research Part C: Emerging
  Technologies}}} \textbf{\bibinfo{volume}{100}}, \bibinfo{pages}{330 -- 353},
  \doiprefix\url{https://doi.org/10.1016/j.trc.2019.01.030}
  (\bibinfo{year}{2019}).

\bibitem{lyon}
\bibinfo{author}{Veve, C.} \& \bibinfo{author}{Chiabaut, N.}
\newblock \bibinfo{journal}{\bibinfo{title}{Estimation of the shared mobility
  demand based on the daily regularity of the urban mobility and the similarity
  of individual trips}}.
\newblock {\emph{\JournalTitle{PLOS ONE}}} \textbf{\bibinfo{volume}{15}},
  \bibinfo{pages}{1--15}, \doiprefix\url{10.1371/journal.pone.0238143}
  (\bibinfo{year}{2020}).

\bibitem{Bettencourt2007}
\bibinfo{author}{Bettencourt, L. M.~A.}, \bibinfo{author}{Lobo, J.},
  \bibinfo{author}{Helbing, D.}, \bibinfo{author}{K{\"{u}}hnert, C.} \&
  \bibinfo{author}{West, G.~B.}
\newblock \bibinfo{journal}{\bibinfo{title}{{Growth, innovation, scaling, and
  the pace of life in cities}}}.
\newblock {\emph{\JournalTitle{Proceedings of the National Academy of
  Sciences}}} \textbf{\bibinfo{volume}{104}}, \bibinfo{pages}{7301--7306},
  \doiprefix\url{10.1073/pnas.0610172104} (\bibinfo{year}{2007}).

\bibitem{Leitao2016}
\bibinfo{author}{Leit{\~{a}}o, J.~C.}, \bibinfo{author}{Miotto, J.~M.},
  \bibinfo{author}{Gerlach, M.} \& \bibinfo{author}{Altmann, E.~G.}
\newblock \bibinfo{journal}{\bibinfo{title}{{Is this scaling nonlinear?}}}
\newblock {\emph{\JournalTitle{Royal Society Open Science}}}
  \textbf{\bibinfo{volume}{3}}, \bibinfo{pages}{150649},
  \doiprefix\url{10.1098/rsos.150649} (\bibinfo{year}{2016}).

\bibitem{Altmann2020}
\bibinfo{author}{Altmann, E.~G.}
\newblock \bibinfo{journal}{\bibinfo{title}{{Spatial interactions in urban
  scaling laws}}}.
\newblock {\emph{\JournalTitle{PLOS ONE}}} \textbf{\bibinfo{volume}{15}},
  \bibinfo{pages}{1--12}, \doiprefix\url{10.1371/journal.pone.0243390}
  (\bibinfo{year}{2020}).

\bibitem{raz1997sub}
\bibinfo{author}{Raz, R.} \& \bibinfo{author}{Safra, S.}
\newblock \bibinfo{title}{A sub-constant error-probability low-degree test, and
  a sub-constant error-probability pcp characterization of np}.
\newblock In \emph{\bibinfo{booktitle}{Proceedings of the Twenty-Ninth Annual
  ACM Symposium on Theory of Computing}}, STOC '97, \bibinfo{pages}{475–484},
  \doiprefix\url{10.1145/258533.258641} (\bibinfo{publisher}{Association for
  Computing Machinery}, \bibinfo{address}{New York, NY, USA},
  \bibinfo{year}{1997}).

\bibitem{Pedersen2020}
\bibinfo{author}{Pedersen, M.~G.} \& \bibinfo{author}{Meneghini, M.}
\newblock \bibinfo{journal}{\bibinfo{title}{A simple method to quantify
  country-specific effects of covid-19 containment measures}}.
\newblock {\emph{\JournalTitle{medRxiv}}}
  \doiprefix\url{10.1101/2020.04.07.20057075} (\bibinfo{year}{2020}).
\newblock
  \eprint{https://www.medrxiv.org/content/early/2020/04/10/2020.04.07.20057075.full.pdf}.

\bibitem{Odagaki2020}
\bibinfo{author}{Odagaki, T.}
\newblock \bibinfo{journal}{\bibinfo{title}{Analysis of the outbreak of
  covid-19 in japan by siqr model}}.
\newblock {\emph{\JournalTitle{Infectious Disease Modelling}}}
  \textbf{\bibinfo{volume}{5}}, \bibinfo{pages}{691 -- 698},
  \doiprefix\url{https://doi.org/10.1016/j.idm.2020.08.013}
  (\bibinfo{year}{2020}).

\bibitem{Hethcote2002}
\bibinfo{author}{Hethcote, H.}, \bibinfo{author}{Zhien, M.} \&
  \bibinfo{author}{Shengbing, L.}
\newblock \bibinfo{journal}{\bibinfo{title}{Effects of quarantine in six
  endemic models for infectious diseases}}.
\newblock {\emph{\JournalTitle{Mathematical Biosciences}}}
  \textbf{\bibinfo{volume}{180}}, \bibinfo{pages}{141 -- 160},
  \doiprefix\url{https://doi.org/10.1016/S0025-5564(02)00111-6}
  (\bibinfo{year}{2002}).

\bibitem{Cao2019}
\bibinfo{author}{Cao, Z.}, \bibinfo{author}{Feng, W.}, \bibinfo{author}{Wen,
  X.}, \bibinfo{author}{Zu, L.} \& \bibinfo{author}{Cheng, M.}
\newblock \bibinfo{journal}{\bibinfo{title}{Dynamics of a stochastic siqr
  epidemic model with standard incidence}}.
\newblock {\emph{\JournalTitle{Physica A: Statistical Mechanics and its
  Applications}}} \textbf{\bibinfo{volume}{527}}, \bibinfo{pages}{121180},
  \doiprefix\url{https://doi.org/10.1016/j.physa.2019.121180}
  (\bibinfo{year}{2019}).

\bibitem{lauer2020incubation}
\bibinfo{author}{Lauer, S.~A.} \emph{et~al.}
\newblock \bibinfo{journal}{\bibinfo{title}{The incubation period of
  coronavirus disease 2019 (covid-19) from publicly reported confirmed cases:
  estimation and application}}.
\newblock {\emph{\JournalTitle{Annals of internal medicine}}}
  \textbf{\bibinfo{volume}{172}}, \bibinfo{pages}{577--582}
  (\bibinfo{year}{2020}).

\end{thebibliography}

\end{document}